\documentclass[%
amsmath,amssymb,
aps,
prl,
floatfix,
]{revtex4-2}
\setcitestyle{super,open={},close={}}

\usepackage{graphicx}
\graphicspath{{figures/}}
\usepackage{dcolumn}
\usepackage{bm}
\usepackage{xcolor}
\usepackage[hidelinks]{hyperref}
\usepackage[mathlines]{lineno}
\usepackage{lipsum}
\usepackage[T1]{fontenc}

\usepackage{glossaries}
\newacronym{HCT}{HCT}{heterodyne cavity-tracking}
\newacronym{FSR}{FSR}{free spectral range}
\newacronym{EOM}{EOM}{electro-optic modulator}
\newacronym{NPRO}{NPRO}{non-planar ring oscillator}
\newacronym{LGWA}{LGWA}{Lunar Gravitational-Wave Antenna}
\newacronym{aLIGO}{aLIGO}{Advanced Laser Interferometer Gravitational-Wave Observatory}
\newacronym{LISA}{LISA}{Laser Interferometer Space Antenna}
\newacronym{KAGRA}{KAGRA}{Kamioka Gravitational Wave Detector}
\newacronym{BOSEM}{BOSEM}{Birmingham Optical Sensor and Electro-Magnetic actuator}
\newacronym{GW}{GW}{gravitational wave}
\newacronym{PDH}{PDH}{Pound-Drever-Hall}
\newacronym{RFSoC}{RFSoC}{radio-frequency system-on-chip}
\newacronym{ULE}{ULE}{Ultra-Low Expansion}
\newacronym{RF}{RF}{radio-frequency}
\newacronym{PID}{PID}{proportional-integral-derivative}
\newacronym{ASD}{ASD}{amplitude spectral density}
\newacronym{d.o.f.}{d.o.f.}{degree of freedom}
\newacronym{LPSD}{LPSD}{logarithmic power spectral density}

\usepackage{siunitx}
\usepackage{amsmath}
\usepackage{orcidlink}
\usepackage{makecell}
\usepackage{tablefootnote}
\definecolor{orcidlogocol}{HTML}{A6CE39}

\begin{document}

\makeatletter
\begin{center}
  {\huge\bfseries Demonstration of a compact optical resonator-based displacement sensing technique with sub-femtometer precision \par\vspace{1.5em}}
  {\normalsize
    \textbf{Shreevathsa Chalathadka Subrahmanya} \orcidlink{0000-0002-9207-4669}$^{*}$\\
    Institute of Experimental Physics, University of Hamburg, Luruper Chaussee 149, 22761 Hamburg, Germany \\
    \href{mailto:shreevathsa.subrahmanya@uni-hamburg.de}{shreevathsa.subrahmanya@uni-hamburg.de}\\
    Telephone: \href{tel:+4940239526271}{+49-40-2395-26271}
    \par\vspace{1em}
    
    \textbf{Jonathan Joseph Carter} \orcidlink{0000-0001-8845-0900}\\
    Max Planck Institute for Gravitational Physics (Albert Einstein Institute), Callinstr. 38, 30167 Hannover, Germany \\
    Institute for Gravitational Physics of the Leibniz Universit\"at Hannover, Callinstr. 38, 30167 Hannover, Germany\\
    Current address: German Aerospace Center, Institute for Satellite Geodesy and Inertial Sensing, Callinstr. 30b, 30167 Hannover, Germany\\
    \href{mailto:jonathan.carter@dlr.de}{jonathan.carter@dlr.de}
    \par\vspace{1em}
    
    \textbf{Oliver Gerberding} \orcidlink{0000-0001-7740-2698}\\
    Institute of Experimental Physics, University of Hamburg, Luruper Chaussee 149, 22761 Hamburg, Germany \\
    \href{mailto:oliver.gerberding@uni-hamburg.de}{oliver.gerberding@uni-hamburg.de}
    \par\vspace{1em}
  }
\end{center}
\vspace{1em}
\makeatother

\title{}
\date{\today}

\begin{abstract}
    We demonstrate sub-femtometer displacement-sensing results achieved with a compact optical resonator-based laser interferometry technique called heterodyne cavity-tracking, intended for local displacement or inertial sensing with ultra-high sensitivity.
    Displacement sensing at this sensitivity is required for ambitious improvements to current gravitational-wave detectors and to enable future ground- and space-based observatories. 
    The optical topology employs a centimeter-scale dynamic cavity incorporating a proof mass, and the relative length fluctuations of this cavity are measured using a heterodyne readout. 
    The fundamental limits of the technique lie significantly below the femtometer level and are ultimately defined by the coating thermal noise of the cavity mirrors. 
    In our experimental demonstration, we achieve a sub-femtometer per Hz$^{1/2}$ displacement sensitivity for Fourier frequencies above \qty{8}{\hertz} and a sub-picometer per Hz$^{1/2}$ sensitivity above \qty{3}{\milli\hertz}, with the sensitivity at lower frequencies limited by mechanical and temperature-induced noise sources. 
    When the length of the dynamic cavity was intentionally actuated, the technique could track a maximum motion of about \qty{0.6}{\micro\meter}, thereby achieving a dynamic range of roughly ten orders of magnitude in displacement sensing.
    We thus demonstrate the key features of this scheme---sub-femtometer performance and a dynamic range spanning ten orders of magnitude---in a laboratory setting, paving the way for development of an integrated system. 
    Such a system is a currently unrealized technology that is necessary for precision physics experiments in the coming decades.
\end{abstract}

\maketitle 

\section{Introduction}
\label{sec:intro}

Local displacement and inertial sensing are essential components of many high-precision experiments, including---but not limited to---gravitational-wave detectors.
When it comes to the detection of \glspl{GW} on Earth, it is necessary to keep the test masses decoupled from intrinsic ground motion, which is achieved by suspending them from a multi-stage pendulum, often combined with an actively stabilized platform~\cite{Matichard2015a,Matichard2015b}. 
The change in position of the platform or of the stages in the passive suspensions is monitored by several displacement and inertial sensors, and this information is then used in a closed-loop control system to maintain active stabilization. 
Existing observatories such as the \gls{aLIGO}~\cite{Capote2025}, Advanced Virgo~\cite{Acernese2014}, \gls{KAGRA}~\cite{Akutsu2020}, GEO600~\cite{Dooley2016}, as well as future detectors like the Einstein Telescope~\cite{Punturo2010} and the Cosmic Explorer~\cite{Abbott2017a}, all benefit from such sensors.

Current displacement sensors are primarily based on the concept of shadow sensors, with a prominent variant known as \glspl{BOSEM}~\cite{Cooper2023,Akutsu2020a}.
Standard \glspl{BOSEM} exhibit a sensor noise level of approximately \qty{0.3}{\nano\meter/\sqrt{Hz}} within the \qtyrange[range-units=single,range-phrase=--]{1}{10}{\hertz} band~\cite{Cooper2023}.
To achieve higher displacement sensitivity and, consequently, improved inertial measurements, laser interferometric options are being explored.
One such candidate is a homodyne quadrature-based interferometer that has demonstrated a sensitivity of about \qty{0.1}{\pico\meter/\sqrt{Hz}} within the \qtyrange[range-units=single,range-phrase=--]{1}{10}{\hertz} band, featuring multi-fringe readout capabilities~\cite{Cooper2018}.
Building upon this readout technique, the current state-of-the-art displacement noise performance across this frequency band reaches tens of femtometers~\cite{Carter2025}.
An alternative approach, investigated to realize compact sensors with up to \qty{10}{\femto\meter/\sqrt{\hertz}} sensitivity, involves deep-frequency modulation interferometry~\cite{Isleif2019,Smetana2022,Eckhardt2024,Alvarez2025}.

The sensitivity of our sensor suite limits the level of isolation that can be achieved by the test masses.
At low frequencies, control-loop-induced displacement noise is one of the primary limiting factors for the sensitivity of current ground-based detectors, and this can be improved by employing sensors with higher precision~\cite{vanDongen2023,Weickhardt2025}.
Future detectors, such as the Einstein Telescope, aim to push this lower-frequency bound further, and thus will require significant advancements over current techniques.
To accommodate the sensor head around the test masses, the sensors must be no larger than a few centimeters, which underlines the need for compact sensing schemes for controlling the test-mass suspension stages. 
Together, these considerations motivate continuing development of novel, compact, and high-precision inertial and displacement sensors.

While ongoing improvements and planning for current and future ground-based laser interferometric \gls{GW} detectors continue, the scientific community is also working on realizing space-based detectors targeting lower Fourier frequencies.
One well-known mission is \gls{LISA}~\cite{Danzmann2003}, which is based on a laser interferometric concept.
It is designed to be sensitive to \glspl{GW} in the milli-Hertz band, in contrast to the audio band in which terrestrial observatories operate.
\Gls{LISA} and its precursor mission, \gls{LISA} Pathfinder, employ test mass displacement readout within spacecraft to enable \gls{GW} detection.
It aims to utilize heterodyne interferometry with picometer-precision for its measurements.
To bridge the gap between the above two frequency bands---namely, to be sensitive to \glspl{GW} in the deci-Hertz band---there is a conceptual idea called \gls{LGWA}. 

With the \gls{LGWA}, an array of high-end seismometers will be deployed on the Moon to monitor its normal modes within the frequency band from \qty{1}{\milli\hertz} to \qty{1}{\hertz} (ref.~\citenum{Harms2021}).
This setup would enable the detection of heavier binary black hole inspirals compared to ground-based detectors and provide early warnings for compact binary black hole and binary neutron star systems, among other capabilities.
The \gls{LGWA} mission is still in the early planning phase, with the technological aspects of the payload currently under study and development~\cite{vanHeijningen2023}.
The current plan involves deploying an array of four seismic stations within a permanently shadowed crater at the Moon's South Pole, each containing two femtometer-class inertial sensors.
The proof mass of these sensors is suspended from a frame rigidly attached to the lunar surface. 
The frame follows the Moon's elastic response to passing \glspl{GW}, producing a differential motion between the proof mass and the frame.
Consequently, the \gls{GW} signal is recorded in this differential motion.
Clearly, the detection principle of \gls{LGWA} is quite different from that of laser interferometer-based detectors.
Nonetheless, the ultimate requirement is achieving a sub-femtometer level readout of the proof mass motion in the inertial sensors, and so far, none of the reported readout techniques meet this standard.

Such high-precision displacement readouts are not only essential for \gls{GW} detection but also serve as critical components across a diverse range of scientific, medical, and industrial domains~\cite{Barbour2001}.
Commercial applications of compact displacement sensors already exist in extreme ultraviolet lithography systems~\cite{Hambric2025,Tonoy2025}.
Furthermore, fields such as particle physics, material science, and seismology could greatly benefit from enhanced displacement or inertial sensors~\cite{Caiazza2017,Reinhard2022,Smetana2025}.

Motivated by these requirements, we demonstrate a compact optical resonator-based displacement sensing technique, termed \gls{HCT}, with a sub-femtometer per Hz$^{1/2}$ readout noise floor.
The interferometer concept is based on prior work by Eichholz et al.~\cite{Eichholz2015} and is further detailed in ref.~\cite{ChalathadkaSubrahmanya2025a}, where a proof-of-principle experiment achieved sensitivities ranging from tens to hundreds of femtometers.
\Gls{HCT} relies on measuring the displacement of the proof mass as a change in the optical cavity length.
To accomplish this, a dynamic cavity comprising the proof mass is constructed, and its length change is inferred from the frequency of a laser that is set to follow the cavity resonance.
Since directly tracking the laser frequency---which is on the order of hundreds of terahertz---is currently technically unfeasible, a heterodyne readout method is employed, in which the laser frequency is compared to another frequency reference.
The differential frequency between the two, known as the beat note, contains the information about the displacement of the proof mass relative to the stability of the frequency reference.

\begin{figure}[htbp]
    \centering
    \begin{minipage}{0.75\textwidth}
        \centering
        \includegraphics[width=\textwidth]{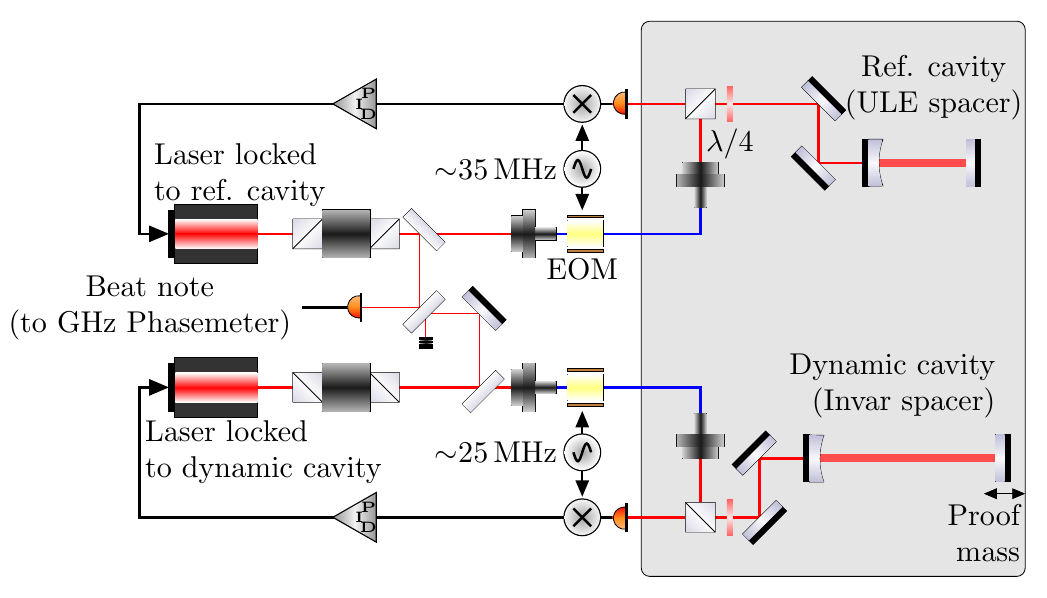}
        (a)\\
    \end{minipage}
    \begin{minipage}{0.5\textwidth}
        \centering
        \vspace{1em}
        \includegraphics[width=\textwidth]{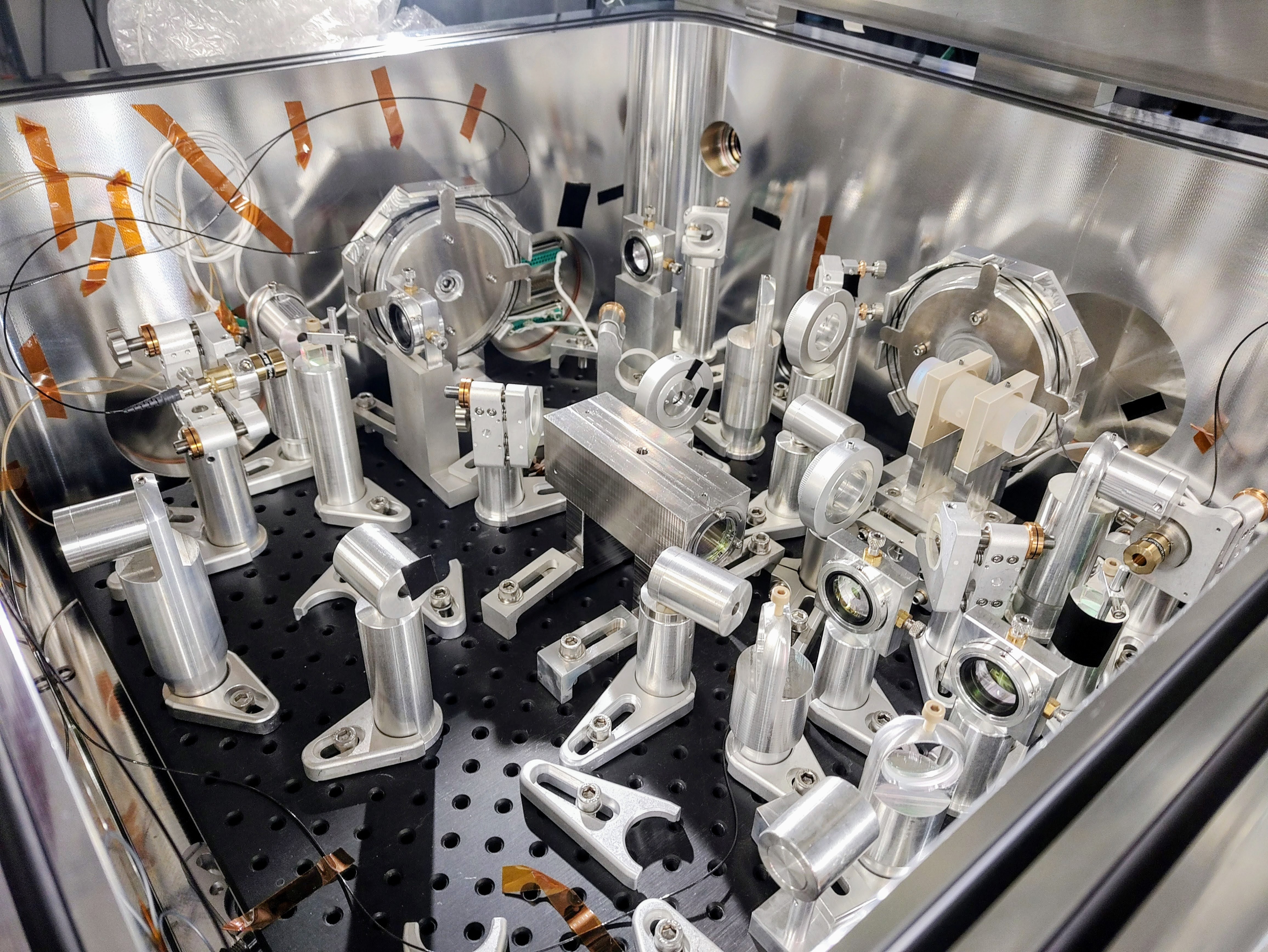}
        (b)\\
    \end{minipage}
    \begin{minipage}{0.95\textwidth}
        \centering
        \vspace{1em}
        \includegraphics[width=\textwidth]{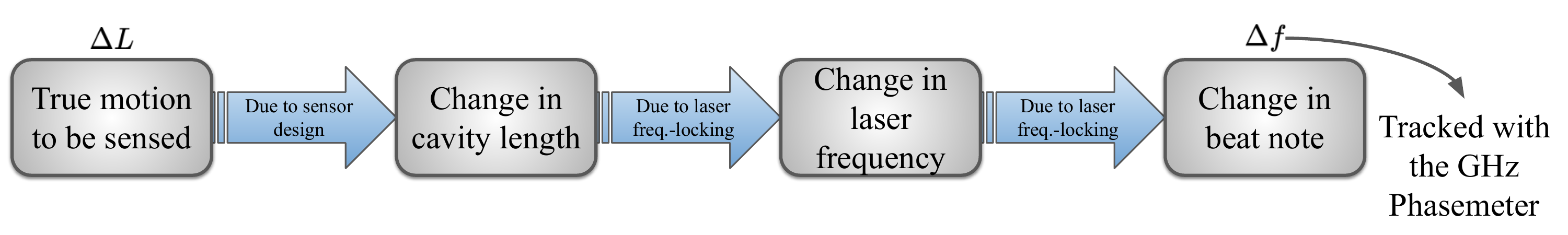}
        (c)\\
    \end{minipage}

    \caption{(a) The experimental layout of the \acrfull{HCT}.
    This optical scheme is chosen for the measurement of achievable displacement sensitivity, where two lasers are locked to the resonances of two optical cavities using the \acrfull{PDH} locking technique.
    The beat note between the two lasers is detected in free-space and it carries the information of interest.
    (b) Photograph of a part of the experimental setup inside the vacuum tank.
    (c) \Gls{HCT} concept shown as a flowchart.}
    \label{fig:huddleLayout}
\end{figure}

Our implementation of the displacement sensor is sketched in FIG.~\ref{fig:huddleLayout}.
One laser is used to follow the resonance of the dynamic cavity, the length of which we are interested in.
To facilitate this, we employ the \gls{PDH} locking scheme~\cite{Drever1983,Black2001}, a well-established method for laser frequency stabilization.
As a frequency reference, we use another similar laser stabilized to a stable reference cavity.
The beat note between the two lasers is read out using a fast photodetector and sent to a frequency-tracking instrument, which we refer to as a phasemeter.
The change in the length of the dynamic cavity manifests as a shift in the beat note frequency, which is retrieved in real-time by the phasemeter readout.

\section{Results}
\label{sec:performance}

We first examine the displacement sensitivity of this readout method using a mechanically fixed stand-in for the \emph{dynamic} cavity.
Subsequently, we deliberately modulate the length of the dynamic cavity using a piezoelectric stack and measure the resulting motion to assess the achievable dynamic range of our displacement sensor.

\subsection{Validation of the displacement sensitivity}

To experimentally demonstrate the achievable displacement sensitivity of the \gls{HCT}, we first measure the length stability of a \qty{94}{\milli\meter}-long optical cavity constructed using Invar, a nickel-iron alloy with a very low coefficient of thermal expansion, as the spacer material.
This probe cavity is an over-coupled cavity in a convex-planar configuration.
The in-coupling mirror has a power reflectivity of \qty{99}{\percent} and a radius of curvature of \qty{0.5}{\meter}, whereas the end mirror of the probe cavity is flat with a reflectivity slightly above \qty{99}{\percent}.
The high-reflectivity mirrors are held in place using metallic blade springs, the outer frame of which is screwed to the Invar spacer.
As a reference, we employ another stable cavity of length \qty{50}{\milli\meter}, also an over-coupled cavity in the same configuration.
This time, the in-coupling mirror has a reflectivity of \qty{98.5}{\percent} and a radius of curvature of \qty{0.25}{\meter}, and the flat end mirror again has a reflectivity slightly above \qty{99}{\percent}.
This reference cavity is constructed by gluing high-reflectivity mirrors onto the annuli of a cylindrical \gls{ULE} glass spacer with a hole bored through the middle.

To lock the primary laser to one of the resonances of the probe cavity and the other laser to that of the reference cavity, any of the available laser frequency stabilization methods can be employed.
A previous demonstration utilized a heterodyne stabilization technique for this purpose~\cite{ChalathadkaSubrahmanya2025a}.
However, the involved digital demodulation and the challenges in maintaining the required demodulation phase pose significant limitations in probing the actual displacement sensitivity of the \gls{HCT} scheme~\cite{Beck2025,chalathadka_subrahmanya_shreevathsa_2025_18337}.
Therefore, the \gls{PDH} locking technique~\cite{Drever1983,Black2001} was chosen for both lasers.

Once both lasers are locked to a resonance of their respective cavities, the beat note between them is tracked using the GHz Phasemeter~\cite{ChalathadkaSubrahmanya2025}.
The relative change in the primary laser frequency is equivalent to the relative change in the length of the dynamic cavity.
Assuming that the change in the laser frequency is the same as the change in the beat note, this effectively measures the relative displacement of the probe cavity with respect to the stability of the reference cavity. 
Since both cavities are positioned next to each other inside the same vacuum tank, one can expect some level of common-mode noise (vibrations and temperature variations) that would not be sensed in this differential readout.
This should facilitate the detection of the achievable noise floor of the \gls{HCT} scheme.

\begin{figure}
  \centering
  \includegraphics[width=.75\linewidth]{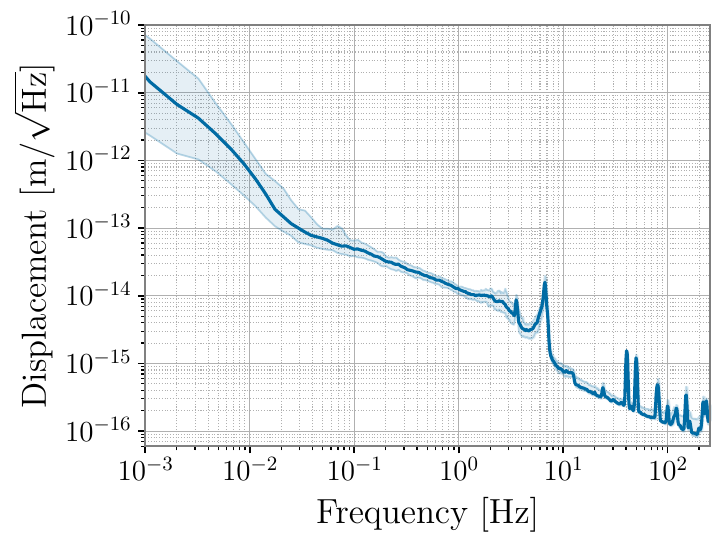}
  \caption{Demonstrated displacement noise of the probe cavity.
  The solid blue trace represents the median value of the spectrogram and corresponds to the \acrfull{ASD} of the displacement.
  A 95th and a 5th percentile confidence bands are also evaluated and shown in the plot.
  This prototype sensing setup has achieved sub-femtometer precision in displacement measurements for Fourier frequencies above \qty{8}{\hertz}.}
  \label{fig:result_ASD}
\end{figure}

FIG.~\ref{fig:result_ASD} shows the displacement sensitivity achieved with our experimental setup.
This result was obtained from a long-term measurement spanning over the weekend, totaling \qty{46}{\hour}.
The recorded beat note time series was divided into multiple segments, each with \qty{30}{\minute} duration.
These data segments were curated to remove glitches or anomalies in the recorded data.
The anomalies were identified to originate from \gls{RF} cross-talk or similar behavior in the frequency readout stage, as the corresponding anomalies were not present in the laser actuation signals.
Accordingly, 15 data segments (corresponding to a total measurement time of \qty{7.5}{\hour}) were discarded from the analysis. 
These frequency time series can be converted into displacement time series by multiplying by the length of the probe cavity and dividing by the laser's absolute frequency (approximately \qty{281.95}{\tera\hertz}, corresponding to a wavelength of \qty{1064}{\nano\meter}).

The displacement time series was initially detrended to remove any linear trend, and then the spectrogram was calculated using the \gls{LPSD} estimation technique~\cite{Trbs2006}.
The logarithmic frequency axis provides results that are more informative than standard spectrograms.
Leaving the linear drift in the time segments would introduce spectral artifacts at the low-frequency end.
The square root of the spectrogram has units of \unit{\meter/\sqrt{\hertz}}, which we use as a metric to quantitatively analyze the performance.
In other words, each line of such a spectrogram corresponds to an \gls{ASD} computed with a custom averaging method.
Furthermore, we determine the median by taking the 50th percentile of the spectrogram, which is represented by the solid blue trace in FIG.~\ref{fig:result_ASD}.
This approach also allows straightforward calculation of confidence bands for the particular measurement.
In FIG.~\ref{fig:result_ASD}, we show the confidence bands corresponding to the 95th and 5th percentiles.
For the data analysis, we used the Python package \verb|Spicypy|~\cite{spicypy}.

It is evident from the plot that we have demonstrated a displacement sensing precision of sub-femtometers for Fourier frequencies above \qty{8}{\hertz} using the \gls{HCT} scheme.
The convergence of the upper and lower confidence bands near the median confirms the reliability of the measured sensitivity.
The increase in uncertainty at frequencies below \qty{0.1}{\hertz} is attributed to non-stationary noise, predominantly residual temperature noise coupling into the cavity length.

\subsection{Noise analysis}
\label{sec:noise}

Before analyzing the noise sources that limit the current experimental demonstration, a brief discussion of the fundamental readout noise floor of the \gls{HCT} is considered.
In this readout technique, noise can couple into the system at two points: during the laser locking to the dynamic cavity and during the measurement of the beat frequency.
This analysis assumes that the frequency reference used to generate the beat note is an ideal system and, therefore, does not contribute to the noise budget.
Noise sources affecting the frequency readout are negligible, and a realistic noise budget lies three orders of magnitude below the femtometer regime~\cite{chalathadka_subrahmanya_shreevathsa_2025_18337}.
In contrast, noise sources impacting the laser locking loop are critical in determining the sensor's performance.
Given that other major noise contributions---such as residual temperature noise coupling into the cavity length---are suppressed to sufficiently low levels, the \gls{HCT} can, in principle, achieve displacement sensitivities fundamentally limited by the thermal noise of the mirror coatings of the dynamic cavity.
This enables the \gls{HCT} technique to reach sub-femtometer displacement readout noise.

\begin{figure}
    \centering
    \includegraphics[width=.75\linewidth]{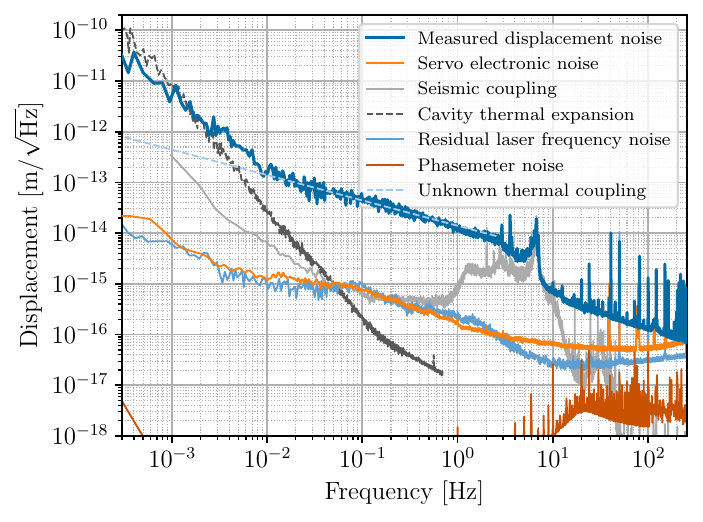}
    \caption{The \acrfull{ASD} plot of the measured displacement sensitivity along with the major noise sources.
    The current performance of our sensor is limited at high frequencies by electronic noise in the control loop. 
    Around \qty{7}{\hertz}, residual seismic noise coupling to the system restricts the achievable sensitivity.
    At very low frequencies, temperature-induced coupling to the cavity length results in non-stationary behavior in our experimental demonstration.}
    \label{fig:noisebudget}
\end{figure}

Instead of a purely theoretical noise budget analysis, we aim to quantitatively measure the contributions of different noise sources, with the results presented in FIG.~\ref{fig:noisebudget}.
For the measured displacement noise, a quiet, glitch-free, \qty{5.5}{\hour}-long stretch of data from the long-term measurement presented above was selected.
In contrast to the previous data processing method, here the \gls{ASD} of the selected data is computed using the Welch method~\cite{Welch1967}.

Initially, the noise contribution from the electronics used in the feedback loop was measured in units of \unit{V/\sqrt{\hertz}}.
This voltage signal was converted into units of meters using calibration factors determined by sweeping the laser frequency near the cavity resonance.
When the cavity is scanned across its resonance, the voltage required to shift from one sideband to the other can be measured.
This voltage corresponds to a known frequency shift---specifically, twice the modulation frequency.
With knowledge of the cavity length, this can be straightforwardly converted into an equivalent mirror displacement.
From such measurements, we find that the peaks at the high-frequency end of the spectrum shown in FIG.~\ref{fig:result_ASD} are primarily due to this electronics noise, most likely arising from ground loops caused by the connection of the control loop to the lasers.
Additionally, above \qty{200}{\hertz}, our displacement sensing is limited by this noise source.

During our measurement campaign, we placed three Sercel L-4C geophones on the optical table to measure residual ground vibrations coupling into the experimental setup along three orthogonal axes.
The signal from the geophone aligned with the axis of the probe cavity was scaled using the mechanical transfer function of the seismic isolation chain, which was well characterized in previous experiments~\cite{Carter2024b}. 
The coupling to the cavity was modeled as that of a simple harmonic oscillator with a resonance of approximately \qty{400}{\hertz}.
This value was estimated based on the resonance of the blade spring mounts used to attach the mirrors to the cavity, as observed in prior measurements~\cite{Carter2020a,Carter2024b}.
Subsequently, the coherence between the seismic signals from the geophones and the displacement signal from our sensor can be analyzed, and the known true motion can be removed using a coherent subtraction routine as described in ref.~\cite{Allen1999,Kirchhoff2017}.
The solid blue trace in FIG.~\ref{fig:noisecomp} is obtained following this analysis.

At the low-frequency end, we suspect that the performance was limited by residual temperature noise.
Recall that the probe and reference cavities are made of different spacer materials; therefore, the length changes due to temperature fluctuations will not be perfectly common-mode.
The presence of such non-stationary, low-frequency noise is evident by the fact that the uncertainty in the measured sensitivity is highest in this region (see FIG.~\ref{fig:result_ASD}).
To quantitatively analyze the temperature coupling, we used in-house developed temperature sensors (similar to those used in ref.~\cite{Beck2025}) placed inside the vacuum tank.
The measured temperature noise was propagated through a modeled transfer function to estimate the corresponding cavity length noise.
This model consisted of a low-pass filter with a corner frequency corresponding to a two-day time constant.
Such a model is consistent with the literature, for example, ref.~\cite{Alvarez2019}. 

Laser frequency noise itself is not a limiting factor for the performance of the \gls{HCT}, as it involves active control of the laser frequency.
For completeness, the residual frequency noise, calculated using the gain of the closed-loop control, is shown in FIG.~\ref{fig:noisebudget}.
It is essential to ensure that the control-loop gain and bandwidth are sufficiently high to not limit the displacement sensitivity.

Although it is known that the noise from the frequency-tracking system is negligible, we performed a sanity check by measuring the overall phasemeter noise.
For this, the superposed laser beams were split into two paths and detected by two identical photodetectors.
The beat notes from both detectors were then tracked using separate channels of the phasemeter.
The measurement noise floor was obtained using the standard split-measurement technique (also known as zero-measurement), where the frequency values from the two channels were subtracted, and the \gls{ASD} of the residuals was calculated.
As expected, the contribution of the phasemeter noise is well below the measured sensitivity of the displacement sensor.

In the mid-frequency range (\qtyrange[range-units=single,range-phrase=--]{0.02}{2}{\hertz}), a form of $1/\sqrt{f}$-type noise currently limits the sensitivity.
The origin of this noise remains unknown.
However, we speculate it may be related to temperature noise affecting the optical fibers used to deliver laser beams into the vacuum tank.
To support this hypothesis, we include a prediction for this noise as a dashed line in FIG.~\ref{fig:noisebudget}, scaled by \qty{14}{\femto\meter}.
Further evidence suggesting a temperature-related origin is the correlation with lab air conditioning: periods of good or bad air conditioning corresponded to lower or higher noise levels.
However, no direct coherence between the temperature probes and the measured length change could be observed.

Scattered light and unwanted secondary reflections from the optical components are also significant noise sources as the measurement approaches femtometer sensitivity.
We have attempted to mitigate these effects by employing optical black light-absorbing material inside the vacuum tank.
Nonetheless, the current performance may still be limited by scattered light, whose nonlinear coupling makes it challenging to model and precisely quantify its contribution.

\subsection{Validation of the operating range}

All displacement sensors have inherent limits on the range of motion they can measure.
For interferometric sensors without feedback, this range can be quite substantial~\cite{Cooper2018}.
However, the noise floor of these sensors can significantly increase when tracking motions on the scale of interferometric fringes~\cite{Smetana2023,Lehmann2025}.
Closed-loop displacement sensors, on the other hand, have a maximum operating range determined by their design, particularly where the actuators reach saturation.
It is crucial for the application that this maximum range exceeds the anticipated motion.

In our sensing scheme, the primary limitations are driven by two factors: the signal bandwidth of the phasemeter and the range of the control loops used to keep the laser frequency resonant with the dynamic cavity.
In our implementation, the \gls{NPRO} laser was equipped with a fast PZT (lead zirconate titanate; a material exhibiting a remarkable piezoelectric effect) controller capable of tuning the laser frequency by approximately \qty{100}{MHz} (about one-tenth of the \gls{FSR} of the cavity), with control bandwidth up to \unit{\kilo\hertz}.
Additionally, a slow temperature controller provided a tuning range of about \qty{3}{\giga\hertz} (roughly one \gls{FSR}) and a maximum unity gain frequency of \qty{0.1}{\hertz}.
Using these, motions over one fringe could be tracked.
However, it is important to note that these are not the fundamental limits, as other cavity control methods have demonstrated significantly higher tuning ranges, such as
widely tunable laser diodes~\cite{Zhang2022a}, fast modulation of sidebands using \glspl{EOM}~\cite{Palmer2022}, or PZT stacks used to control the cavity length itself~\cite{Carter2024b}.

The phasemeter we use has an unparalleled detection bandwidth of \qty{2.048}{\giga\hertz} (ref.~\citenum{ChalathadkaSubrahmanya2025}), which sets the ultimate limit for the operating range of the sensing scheme.
Since the displacement $d$ is related to the frequency shifts of the cavity resonance condition $f$ by 
\begin{equation}
    d = \frac{\lambda}{2}\cdot \frac{f}{\rm{\gls{FSR}}},
\end{equation}
where $\lambda$ is the laser wavelength, one can decrease the frequency shifts due to motion by reducing the cavity \gls{FSR} (i.e., by increasing the cavity length).
However, this approach introduces more stringent requirements on frequency noise, so that it does not become a limiting noise source.

\begin{figure}
    \centering
    \includegraphics[width=.75\linewidth]{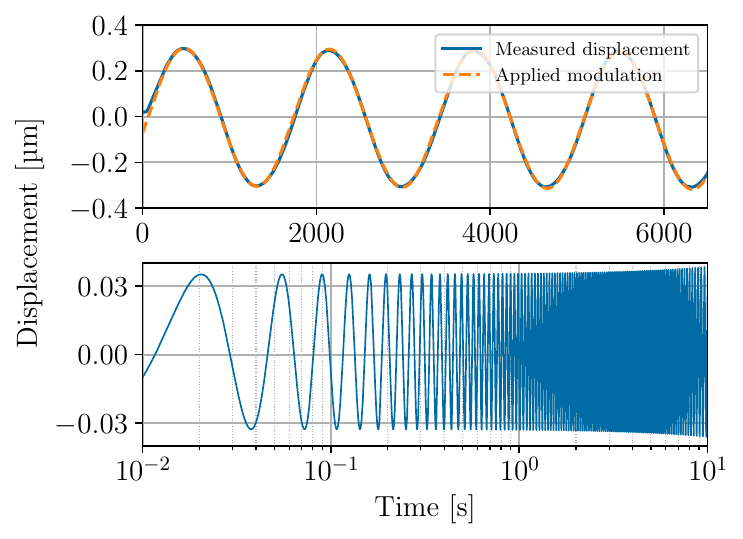}
    \caption{Demonstrating the maximum operating range of our displacement sensor.
    The length of the dynamic cavity was intentionally modulated by applying a sinusoidal signal to the attached PZT stack.
    With a slow modulation, the readout was capable of operating over an interferometric fringe-scale, reaching a maximum total displacement of approximately \qty{0.6}{\micro\meter}, as shown in the top subplot.
    The bottom subplot confirms that the sensor can also handle rapid motions of the dynamic cavity.}
    \label{fig:rangetesting}
\end{figure}

To experimentally demonstrate the achievable operating range, a ring-shaped PZT stack was placed between the Invar spacer and the end mirror of the probe cavity.
By applying an electrical signal to the PZT stack, we can control and vary the length of the probe cavity.
FIG.~\ref{fig:rangetesting} shows the results from two such tests.
Initially, the cavity length was modulated using a low-frequency sine wave.
When the cavity length was changed by about \qty{0.6}{\micro\meter}, the beat frequency between the lasers shifted by approximately \qty{1.8}{\giga\hertz}, and the high-bandwidth phasemeter accurately tracked this change.
Combined with the sensor's sub-femtometer sensitivity, this corresponds to a total dynamic range of roughly ten orders of magnitude for displacement sensing.

To verify whether the readout can keep pace with rapid changes, the cavity length was modulated while sweeping the modulation frequency from \qty{100}{\hertz} down to \qty{1}{\hertz} over a span of \qty{10}{\second}.
To maintain the laser lock with the resonance of the dynamic cavity, the modulation depth was reduced to produce a total displacement of approximately \qty{70}{\nano\meter}.
The results are as expected; the increase in displacement observed at lower modulation frequencies is simply due to the transfer function of the ring PZT.

\subsection{Quantifying the nonlinear behavior}

When a displacement sensor measures the motion of a test mass, the actual motion is not necessarily converted linearly into the readout signal.
Several effects can cause such behavior, including hysteresis, non-stationarity, and bi-directionality~\cite{Stone1998,Smetana2023}.
These effects typically become more pronounced when the measured motion approaches the maximum range of the sensor, when the motion is fast, or when the test mass has multiple degrees of freedom.
Interferometric displacement sensors have been shown to exhibit significant nonlinear behavior, particularly when schemes to suppress laser-noise coupling to displacement noise are used~\cite{Hou1992,Lehmann2025}.

The scheme we present here is, in principle, resistant to nonlinear behavior despite its complexity.
Because we do not add any signals but instead compare two operating points, we observe the actual resulting operating points rather than the signal used to reach them.
This means that we obtain a measurement of the operating point that is determined solely by the gains of the control loops and the response of the phasemeter.
However, we remain subject to nonlinearities that cause the cavity length to change in a non-ideal manner, such as geometric misalignment of the mirrors, which is governed by the planarity of the two spacer faces and the optical alignment.
We therefore aim to estimate the magnitude of this effect.

Nonlinear behavior encompasses a wide range of effects that vary depending on the experimental setup, including their scaling, prominence, and coupling mechanisms.
It is therefore not possible to define a single metric or test that enables a fair comparison with other sensing techniques.
Previous studies have examined residuals after subtracting the measured motion~\cite{Smetana2023,Hou1992} or have used sensor noise as an estimate of nonlinear behavior~\cite{Lehmann2025}.
Using sensor noise as a monitor of nonlinearity requires an independent reference for the motion, which is not available in our scheme.
We therefore focus on studying the residuals after the motion has been subtracted.

Typically, most methods used to induce motion for a displacement sensor introduce their own nonlinear behavior.
To prevent this from limiting the estimation of the sensor's nonlinearity, we analyze only a fraction of the linear sweep around its central region, where the actuator's nonlinear behavior is minimal.
In our implementation of this test, a ramp of \qty{40}{\nano\meter} was applied, and a central range of \qty{20}{\nano\meter} was selected for analysis.
A slow ramp with a frequency of \qty{0.01}{\hertz} was used, as slow actuation further minimizes the nonlinear response of the PZT actuator.

\begin{figure}
    \centering
    \includegraphics[width=0.75\linewidth]{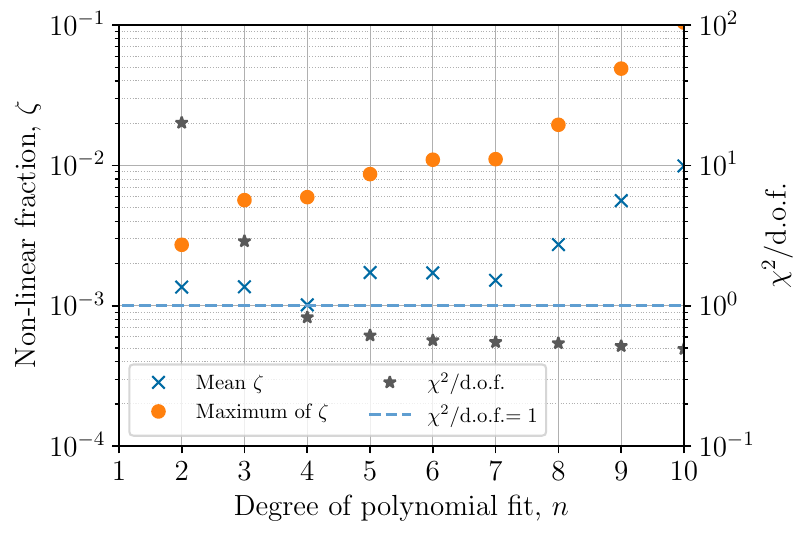}
    \caption{A figure displaying the nonlinear behavior of the sensor.
    The left axis shows the nonlinear behavior of the sensor, $\zeta$, for different degrees of polynomial used to fit the ramp data.
    The right axis shows the $\chi^2$ per \acrfull{d.o.f.} values calculated from these fits, with the optimal value of $\chi^2$ per \gls{d.o.f.}$=1$ highlighted.
    This optimal condition occurs near the fourth-degree polynomial fit.}
    \label{fig:nlintest}
\end{figure}

With the ramp applied and the sensor response measured, the nonlinear contribution can be estimated.
We assess the nonlinear content by fitting an $n$th-order polynomial to the selected region of the ramp.
As a first step, we fit the measured ramp to a linear function
\begin{equation}
    s_{\rm{meas}}=p^t_1t+p^t_2,
\end{equation}
where $s_{\rm{meas}}$ is the measured displacement, $p^t_i$ are the fitted parameters describing the mapping from time to displacement, and $t$ is measurement time.
We define $s=0$ as the center of the scan and set $t=0$ at the same point.
The ideal ramp can then be expressed as
\begin{equation}
    s^{\prime}=p^t_1t.
\end{equation}
The measured deviations from this ideal ramp are then fitted using a least-squares procedure with an $n$th-order polynomial,
\begin{equation}
    s_{\rm{meas}}=\sum_{i=0}^{i=n} p_i(s^{\prime}) ^i,
\end{equation}
where $p_i$ are the fitted polynomial coefficients.

We then define the nonlinear fraction content as
\begin{equation}
    \zeta_n= \dfrac{\sum_{i=2}^{i=n} \mid p_i(s^{\prime}) ^i\mid}{\mid s^{\prime} \mid},
    \label{eqn:nlin}
\end{equation}
which is computed for every data point in the scan. 
Meaningful inferences can then be drawn from these data by evaluating, for example, the mean value over the scan or the maximum deviation within the scan, which typically occurs near the extrema.
These results are shown in FIG.~\ref{fig:nlintest}.

Initially, we observe that the nonlinear contribution increases as higher-order terms are added to the polynomial fit.
However, this increase may also result from overfitting.
To assess this, a $\chi^2$ per \gls{d.o.f.} test was performed.
The variance used for the $\chi^2$ estimate was obtained from data recorded with no applied ramp signal.
We find that the $\chi^2$ per \gls{d.o.f.} is closest to $1$ for a fourth-degree polynomial fit.
We therefore adopt this polynomial degree for our estimate of the nonlinearity.
It remains possible that contributions from other polynomial orders are present in the sensor readout, but these would lie significantly below the noise level in our measurement, which had a variance of \qty{800}{\femto\meter} under these conditions.

The measurements in FIG.~\ref{fig:nlintest} show that, for a motion with a \qty{20}{\nano\meter} scale, the maximum deviation remains below the sub-percent level, with an average value of approximately \qty{0.1}{\percent}.
Both values should be interpreted as upper estimates of the nonlinear behavior at these displacement scales.
The experiment was then repeated by connecting a high-voltage supply to the PZT stack in order to induce a full-fringe displacement.
Under these conditions, the noise introduced by the actuator increased dramatically, allowing us to confirm only that the maximum nonlinearity remained below \qty{3}{\percent} for motions of \qty{200}{\nano\meter}.
This measurement was fully dominated by the PZT response.

\section{Discussion}
\label{sec:application}

The demonstrated displacement sensitivity, which is below one femtometer per Hz$^{1/2}$ for Fourier frequencies above \qty{8}{\hertz}, is very promising for many high-precision application, including the detection of \glspl{GW}.
Achieving a precision of better than \qty{100}{\femto\meter/\sqrt{\hertz}} for frequencies above \qty{30}{\milli\hertz} is rare for compact, centimeter-scale displacement sensors.
The precision reported here is comparable to the exceptionally low residual displacement noise of \qty{32.1}{\femto\meter/\sqrt{\hertz}} achieved---down to much lower frequencies---by the \gls{LISA} Pathfinder mission~\cite{Armano2021}.
It is worth noting, however, that the performance of \gls{LISA} Pathfinder was obtained under conditions of significantly reduced ambient noise and extreme thermal stability, owing to the entire instrument being operated in space. 

\begin{figure}
    \centering
    \includegraphics[width=.75\linewidth]{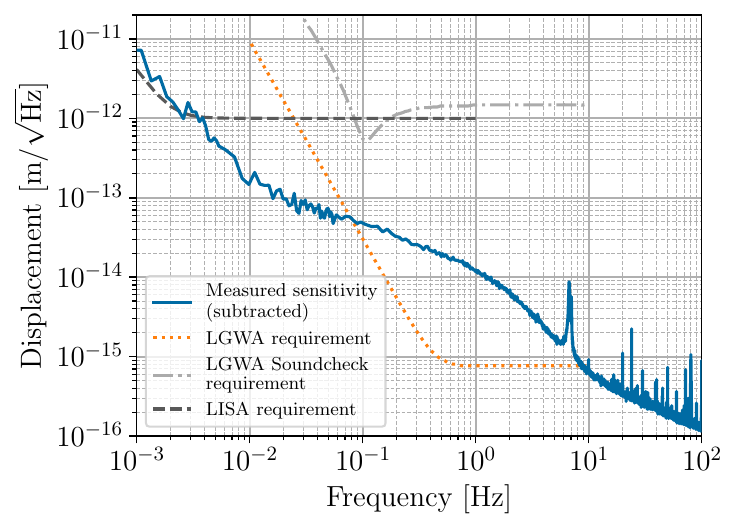}
    \caption{Demonstrated displacement sensitivity of the \acrfull{HCT}, shown together with other high-precision requirements set by space-based \gls{GW} detection missions.
    The solid blue trace represents the measured sensitivity of our sensor after coherent subtraction of the geophone signals.
    The \gls{LGWA} requirement traces are taken from ref.~\cite{Andric2026}.
    (LGWA: Lunar Gravitational-Wave Antenna; LISA: Laser Interferometer Space Antenna.)} 
    \label{fig:noisecomp}
\end{figure}

In FIG.~\ref{fig:noisecomp}, we place our result in the context of two space missions with highly demanding displacement-sensitivity requirements.
The shoulder-like behavior of our spectrum around \qty{1}{\hertz} suggests a likely contribution from stray light, which was masked by residual seismic noise in FIG.~\ref{fig:result_ASD}.   
Nevertheless, the results obtained under laboratory conditions nearly satisfy the stringent picometer-level stability required by the \gls{LISA} mission~\cite{AmaroSeoane2017}.
As explained in the Introduction, enabling \gls{GW} detection on the Moon using the \gls{LGWA} concept requires an inertial sensing scheme with a sub-femtometer noise floor.
While our scheme fully meets the requirements of the \gls{LGWA} Soundcheck, it also demonstrates performance that approaches the sensitivity needed for the \gls{LGWA} itself.
Such a high-precision inertial sensing would also be essential for future space-based \gls{GW} missions beyond \gls{LISA}~\cite{Backer2019}.

In the context of local displacement sensing for ground-based \gls{GW} detectors, the demonstrated precision is approximately 100 to 1000 times better than that of existing sensors or reported ongoing developments.
However, it should be noted that our sensor is sensitive to only a single \gls{d.o.f.}, namely the longitudinal motion along the dynamic cavity length.
To measure motion along any other axis, a separate dynamic cavity would be required, and its length change would need to be read out using another laser while sharing the same frequency reference.
The phasemeter used here provides a sufficient number of channels to sense all six degrees of freedom of a system.
To employ the \gls{HCT} scheme as a readout method for existing inertial or displacement sensors, one must construct a linear cavity consisting of a fixed input mirror and a reflective surface on the moving proof mass.
This construction approach has been demonstrated in fused silica test masses to develop inertial sensors~\cite{Carter2024b}.

The \gls{HCT} scheme is directly applicable as a readout method for several ongoing inertial-sensor and accelerometer developments.
For example, ref.~\cite[Fig.\,1]{Carter2020a} already presents a sensor geometry that matches the configuration of interest here.
Combining such a sensing head with our \gls{HCT} approach would enable a compact, very high-precision inertial sensor.
To further enhance the sensitivity of existing inertial sensors, approaches such as those presented in ref.~\cite{Cooper2022} may also be explored, using the \gls{HCT} scheme as the readout method.

The \gls{HCT} offers several inherent advantages due to its design.
For a given sensor configuration, the displacement information is obtained directly from the beat note using a linear conversion factor.
This enables straightforward integration of the sensing scheme into real-time systems with closed-loop control.
There is no need for post-processing or for reconstructing the displacement through complex algorithms.
Moreover, no nonlinearity‑correction methods were required to achieve the reported performance.  
These points highlight the simplicity and robustness of the readout approach used in our scheme.

The primary motivation for this work was to experimentally demonstrate a displacement-sensing method with a sub-femtometer readout noise floor.
For any displacement sensor, it is beneficial to achieve both high sensitivity and a large operating range, yielding a high dynamic range for displacement measurements.
From the analytical understanding of the \gls{HCT}, it is evident that a coating thermal noise-limited performance is achievable, placing the readout noise floor several orders of magnitude below the femtometer regime.
However, such a high level of precision had not previously been demonstrated in a tabletop experimental setup.
Our experiment, employing a mechanically fixed \emph{dynamic} cavity, has achieved sub-femtometer precision above \qty{8}{\hertz}, with a peak sensitivity better than \qty{e-16}{\meter/\sqrt{\hertz}} at \qty{200}{\hertz}.
We emphasize that this demonstrated performance is not limited by the sensing scheme.
It may be possible to achieve displacement sensitivities below the requirement of \gls{LGWA}, potentially impacting \gls{GW} sensing capabilities with optical readout.

By using a dynamic cavity with a tunable cavity length, we demonstrated that our sensor can operate over a fringe-scale, tracking a maximum motion of approximately \qty{0.6}{\micro\meter}.
Combined with the demonstrated peak sensitivity, this corresponds to a dynamic range of about ten orders of magnitude for displacement sensing.
Such state-of-the-art displacement readout performance is achievable with the \gls{HCT} scheme primarily due to the use of an ultra-fast phasemeter with GHz-level bandwidth. 

The dominant noise sources of the current setup are largely understood.
To achieve coating thermal noise-limited performance---which would satisfy the \gls{LGWA} requirement in the deci-hertz band---the experimental setup could be improved through lower-noise electronics in the feedback loop, enhanced thermal and seismic isolation, and further mitigation of stray light.
To reduce the latter, the use of free-space \gls{EOM}s and detecting cavity reflections inside the vacuum tank are obvious next steps.
Techniques such as tunable coherence for stray-light reduction~\cite{Voigt2025} also appear very promising.
With continued technological advancements, it may become possible to integrate even higher-bandwidth frequency-tracking instruments in the future, enabling multi-fringe displacement readout with exceptional precision over a wide frequency range.

\section{Materials and methods}
\label{sec:concept}

The lasers used in our experimental demonstration are the continuous-wave, solid-state Nd:YAG lasers (One Mephisto-500 and one Mephisto-1000 by Coherent Corp.), operating at \qty{1064}{\nano\meter} in the \gls{NPRO} configuration.
In the previous experimental demonstration~\cite{ChalathadkaSubrahmanya2025a}, sensitivity was significantly affected by the deployment of fiber-optic components. 
To mitigate this, we have opted for a greater extent of free-space propagation.
This approach was expected to improve the experimental setup by reducing the amount of unwanted parasitic beams, which can cause parasitic etalons~\cite{Jennings2017}.

The laser beam from one of the lasers is polarization-cleaned, and a small portion of the beam is picked off to generate the beat note.
The remaining part is fiber-coupled, and phase-modulation sidebands for \gls{PDH} locking are generated using a fiber-based \gls{EOM} (Photline MPX-LN-0.5 by Lucent Technology Ltd.).
The phase-modulated carrier is then coupled into a vacuum tank via a fiber feed-through.
The moderate-sized vacuum tank sits on an active isolation platform (Accurion i4 by Park Systems), and the optical breadboard within the vacuum sits on passive isolation feet (VIBe\texttrademark\ VIB100 by Newport Corp.) to isolate the experimental setup from residual ground vibrations.  
A stable vacuum level of approximately \qty{6e-3}{mbar} was achieved in the tank using an inbuilt backing pump at the Albert Einstein Institute (AEI), Hannover.
This configuration allows the pumps to be significantly spatially separated from the experiment, preventing vibrations from disturbing the measurement.

Inside the vacuum tank, the laser beam is redirected back into free-space propagation and polarization-cleaned once again.
It is then coupled into the \emph{dynamic} cavity after passing through the appropriate mode-matching lenses.
The reflected light from the cavity is separated from the incoming beam using a combination of a polarizing beam-splitter and a quarter-wave plate.
The reflected light then passes through a transparent window of the vacuum tank and is directed onto the in-house-built photodetector located outside the tank.

A similar optical layout is assembled for a second laser serving as the frequency reference.
Two external \gls{RF} signal generators at frequencies approximately \qty{35}{\mega\hertz} and \qty{25}{\mega\hertz} were used to imprint sidebands onto the two laser frequencies, respectively.
To avoid any \gls{RF} cross-talk and harmonic generation, we selected arbitrary frequency values slightly offset from the round values.
The cavity reflection signals, detected using photodetectors, were then demodulated at these specific frequencies using custom-built analog circuitry. 
The resulting error signals are fed into an analog \gls{PID} controller stage, which produces actuation signals for the corresponding laser drivers.
A combination of proportional, integral, and double-integral gains was used to achieve a control loop with a unity gain frequency of approximately \qty{15}{\kilo\hertz}.
Laser frequency actuation was performed by controlling both the signal applied to the piezoelectric element of the laser (fast actuation) and the laser temperature (slow actuation).
It should be noted that, although the two control loops described above are identical, they operate independently of each other.

Both lasers have a pick-off (approximately \qty{5}{\milli\watt} of the total \qtyrange[range-units=single,range-phrase=--]{500}{1000}{\milli\watt} power) located just after the Faraday isolator. 
The pick-off signals interfere in air using a beam-splitter, and the resulting interference pattern is detected with a high-bandwidth free-space InGaAs detector (DET08CL/M by Thorlabs Inc.).
The radio-frequency component, representing the beat note between the two lasers, is amplified and fed into the \gls{RFSoC}-based phasemeter, which has a bandwidth of \qty{2.048}{\giga\hertz}.
The development and characterization of this frequency-tracking instrument are detailed in ref.~\cite{ChalathadkaSubrahmanya2025}.

\section*{Acknowledgments}
SCS and OG acknowledge support from the Deutsches Zentrum f\"ur Luft- und Raumfahrt (DLR), with funding from the Bundesministerium f\"ur Wirtschaft und Klimaschutz under Project 50OQ2302, as well as support from the Deutsche Forschungsgemeinschaft (DFG, German Research Foundation) under Germany's Excellence Strategy--EXC 2121 ``Quantum Universe'' under Grant 390833306. 
JJC acknowledges support from the Glass Technologies for the Einstein Telescope (GT4ET) Kooperations Project.

The authors would like to thank AEI Hannover for making their laboratory facilities available for conducting the experiments.
We thank Artem Basalaev for valuable suggestions regarding the data analysis and Jan Harms for comments during the LIGO P\&P review process.
We also thank Christian Darsow-Fromm for assistance with the GHz Phasemeter, and Fabian Meylahn and Henning Vahlbruch for their support with the photodetector and servo controls.

\section*{Author contributions}
OG conceived and oversaw the project and contributed to the review and finalization of the manuscript.
SCS and JJC contributed equally to the experimental work, data analysis, and preparation of the manuscript.

\section*{Data availability}
The data that support the findings of this study are available from the corresponding author on reasonable request.

\section*{Conflict of interest}
The authors declare no conflict of interest.

\bibliographystyle{naturemag}
\bibliography{references}
\end{document}